\documentclass[12pt]{article}

\catcode`\@=11
\@addtoreset{equation}{section}

\global\arraycolsep=2pt
\usepackage{amsbsy,amssymb,latexsym,amsfonts,amsmath}
\usepackage{graphicx,color}

\allowdisplaybreaks

\begin{document}
\begin{flushright}
\parbox{4.2cm}
{UCB-PTH-10/11}
\end{flushright}

\vspace*{0.7cm}

\begin{center}
{\Large \bf 
Refined Cigar and $\Omega$-deformed Conifold}
\vspace*{2.0cm}\\
{Yu Nakayama}
\end{center}
\vspace*{-0.2cm}
\begin{center}
{\it Berkeley Center for Theoretical Physics, \\ 
University of California, Berkeley, CA 94720, USA }
\vspace{4.8cm}
\end{center}

\begin{abstract} 
Antoniadis et al proposed a relation between the $\Omega$-deformation and refined correlation functions of the topological string theory. We investigate the proposal for the deformed conifold geometry from a non-compact Gepner model approach. The topological string theory on the deformed conifold has a dual description in terms of the $c=1$ non-critical string theory at the self-dual radius, and the $\Omega$-deformation yields the radius deformation. We show that the refined correlation functions computed from the twisted $SL(2,\mathbf{R})/U(1)$ Kazama-Suzuki coset model at level $k=1$ have direct $c=1$ non-critical string theory interpretations. After subtracting the leading singularity to procure the 1PI effective action, we obtain the agreement with the proposal.
\end{abstract}

\thispagestyle{empty} 

\setcounter{page}{0}

\newpage

\section{Introduction} 
In recent studies of $\mathcal{N}=2$ supersymmetric gauge theories (i.e. with 8 supercharges), the $\Omega$-deformation \cite{Nekrasov:2002qd}\cite{Moore:1997dj}\cite{Losev:1997tp} has been playing a significant role. While the $\Omega$-deformation enables us to exactly compute various quantities in gauge theories, many physically interesting observables (e.g. low energy prepotential) are not affected by the deformation. This has generated novel approaches to attack the non-perturbative aspects of the $\mathcal{N}=2$ supersymmetric gauge theories \cite{Alday:2009aq}\cite{Nekrasov:2009rc}\cite{Dijkgraaf:2009pc}\cite{Nekrasov:2010ka}.

The $\Omega$-deformation in the four-dimensional Euclidean space has  two deformation parameters i.e. $\epsilon_1$ and $\epsilon_2$. It has been long known that when $\epsilon_1 = - \epsilon_2$, the $\Omega$-deformation can be interpreted as higher genus corrections to the topological string partition function \cite{Bershadsky:1993cx}\cite{Antoniadis:1993ze} when the geometric engineering gives the $\mathcal{N}=2$ supersymmetric gauge theory under consideration \cite{Klemm:1996bj}\cite{Katz:1997eq}. In particular, the topological vertex method \cite{Aganagic:2003db} has given a very convenient way to compute the higher genus topological string partition function, and the result completely agrees with the $\epsilon_1 = -\epsilon_2$ gauge theory computation.

It is then natural to attempt to generalize, or refine, the topological string computation so that we recover the two-parameter deformations of the $\mathcal{N}=2$ supersymmetric gauge theories. A heuristic approach to modify the rules of the topological vertex method (known as ``refined topological vertex") has been pursued in \cite{Iqbal:2007ii}, while the physical derivation is still lacking. On the other hand, certain higher derivative F-terms in the topological string theory have been computed in \cite{Antoniadis:2010iq}, and it is proposed that they correspond to the second parameter in the $\Omega$-deformation.

An indirect check of the proposal has been done in the simplest non-trivial $\mathcal{N}=2$ theory based on the deformed conifold through the duality to the heterotic string theory \cite{Morales:1996bp}. The partition function of the topological string theory on the deformed conifold is given by the partition function of the $c=1$ non-critical string theory at the self-dual radius \cite{Ghoshal:1995wm}. 
It is known that the second parameter in the $\Omega$-deformation of the deformed conifold corresponds to the change of the radius from the $c=1$ non-critical string theory viewpoint \cite{Dijkgraaf:2009pc}. The heterotic string one-loop computation has revealed that this is indeed the case. The aim of this paper is to investigate the correspondence directly in the type II string theory.

For this purpose, in section 2, we first review the triality relations among the topological string theory on the deformed conifold, the $c=1$ non-critical string theory at the self-dual radius, and the topologically twisted $SL(2,\mathbf{R})/U(1)$ Kazama-Suzuki coset model \cite{Kazama:1988qp} at level $k=1$ \cite{Mukhi:1993zb}\cite{Ghoshal:1995wm}\cite{Ashok:2005xc}. It is the last coset construction that gives us interpretations of the refined topological amplitudes in terms of the $c=1$ non-critical string theory in section 3. The coset theory is known as a cigar model because the target space shows a geometry of the cigar \cite{Dijkgraaf:1991ba}. The deformed conifold theory realized as a $SL(2,\mathbf{R})/U(1)$ coset theory is a non-compact version of the Gepner-model construction of the Calabi-Yau theories. After subtracting the leading singularity to procure the 1PI effective action, we show that the refinement in the topological amplitudes can be regarded as a radius change deformation of the $c=1$ non-critical string theory as proposed.

\section{Topological cigar, conifold, $c=1$ string triality}
The relation between the topologically twisted $SL(2,\mathbf{R})_k/U(1)$ Kazama-Suzuki coset model \cite{Kazama:1988qp} at level $k=1$ and the $c=1$ non-critical string theory at the self-dual radius was first advocated in \cite{Mukhi:1993zb}. The original construction was based on the Wakimoto representation of the $SL(2,\mathbf{R})$ current algebra, but later in more recent years, it was shown \cite{Ashok:2005xc} that without referring to the specific Wakimoto representation, one can complete the analysis based on the restricted Hilbert space (i.e.  with the so-called improved unitarity bound \cite{Maldacena:2000hw}) by using the spectral flow, which directly appears in the partition function of the $SL(2,\mathbf{R})_1/U(1)$ coset theory.

\subsection{Coset construction}

Let us begin with the parent supersymmetric $SL(2,\mathbf{R})_k$ current algebra generated by $(J^a, \psi^a)$ with the OPE\footnote{In the following, we follow the convention in \cite{Ashok:2005xc} except that our Liouville field $\phi$ will have a flipped sign compared to theirs.}
\begin{align}
J^a(z) J^b(w) &\sim \frac{k g^{ab}}{2(z-w)^2} + \frac{f^{ab}_{c} J^{c}}{z-w} \cr
J^{a}(z) \psi^b(w) &\sim \frac{i f^{ab}_{c} \psi^c}{z-w} \cr
\psi^a(z) \psi^b(w) & \sim \frac{g^{ab}}{z-w} \ ,
\end{align}
where $g_{ab} = \text{diag}(1,1,-1)$ and $f^{123} = 1$. We introduce the bosonic current $j^a$ with level $k+2$ as
\begin{align}
j^a = J^a + \frac{i}{2}f^{a}_{bc} \psi^b \psi^c \ ,
\end{align}
which commutes with the free fermions $(\psi^a, \bar{\psi}^a)$.

The $SL(2,\mathbf{R})/U(1)$ Kazama-Suzuki construction gives the $\mathcal{N} =2$ currents:
\begin{align}
T &= T_{SL(2,\mathbf{R})} - T_{U(1)} \cr
G^{\pm} &= \sqrt{\frac{2}{k}} \psi^{\pm} j^{\mp} \cr
J^R &= \frac{2}{k} J^3 + \psi^+ \psi^- \ ,
\end{align} 
where $T_{U(1)} = -\frac{1}{k}J^3J^3 + \frac{1}{2}\psi^3\partial \psi^3$. The central charge of the Kazama-Suzuki coset is $c = 3 + \frac{6}{k}$. In the following, we will set $k=1$ so that we obtain $c=9$.

To obtain the geometric picture of the coset, it is useful to introduce the gauging of the $U(1)$ symmetry by introducing an additional gauge field \cite{Dijkgraaf:1991ba}. In the axial gauging, we add an extra $\mathcal{N}=1$ multiplet $(X,\psi^X)$ and introduce the BRST charge
\begin{align}
J_{\text{BRST}} = C(J^3 + i\sqrt{\frac{1}{2}}\partial X) + \gamma'(\psi^3 + \psi^X) \ , 
\end{align}
where $(B,C)$ is a $(1,0)$ ghost with the central charge $-2$, and $(\beta',\gamma')$ is a fermionic ghost to remove $\psi^3,\psi^X$ from the spectrum to obtain two free fermions $\psi^{\pm}$ on the cigar. The bosonic gauging currents are
\begin{align}
J_g = J^3 + i\sqrt{\frac{1}{2}}\partial X \ .
\end{align}

For the best comparison to the $c=1$ string theory, it is useful to introduce the Wakimoto representation of the $SL(2,\mathbf{R})$ algebra at level $k=1$:
\begin{align}
j^+ & = \beta \ &  \bar{j}^+ &= -\bar{\beta} \bar{\gamma}^2 + \sqrt{2} \bar{\gamma}\bar{\partial}\phi - 3 \bar{\partial}\bar{\gamma} \cr
j^3 &= - \beta \gamma + \sqrt{\frac{1}{2}} \partial \phi  \ & \bar{j}^3 &= \bar{\beta}\bar{\gamma} - \sqrt{\frac{1}{2}}\bar{\partial}\phi \cr
j^- &= \beta \gamma^2 - \sqrt{2}\gamma\partial\phi + 3\partial \gamma \ & \bar{j}^- &= -\bar{\beta} \ .
\end{align}
With these variables, the energy momentum tensor is given by
\begin{align}
T_{SL(2,\mathbf{R})} = \beta \partial \gamma - \frac{1}{2}(\partial\phi)^2 +\sqrt{\frac{1}{2}} \partial^2 \phi - \frac{1}{2}\psi^+ \partial \psi^- - \frac{1}{2} \psi^- \partial \psi^+ \ .
\end{align}

To connect to the $c=1$ string theory, we perform the topological A-twist: 
\begin{align}
T_{\text{top}} = T +\frac{1}{2} \partial J^R  \ , \ \ \ \bar{T}_{\text{top}} = \bar{T} - \frac{1}{2} \partial \bar{J}^R \ .
\end{align}
In Wakimoto variables they are
\begin{align}
T_{\text{top}} &= - \partial \beta \gamma -\frac{1}{2}(\partial\phi)^2 + \sqrt{2}\partial^2\phi -\frac{1}{2}(\partial X)^2 - \frac{1}{2}\psi^+\partial \psi^- -\frac{1}{2} \psi^-\partial \psi^+ + \frac{3}{2}\partial(\psi^+\psi^-) \ \cr
\bar{T}_{\text{top}} &= - \bar{\partial} \bar{\beta} \bar{\gamma} -\frac{1}{2}(\bar{\partial}\phi)^2 + \sqrt{2}\bar{\partial}^2\phi -\frac{1}{2}(\bar{\partial} X)^2 - \frac{1}{2}\bar{\psi}^+\bar{\partial} \bar{\psi}^- -\frac{1}{2} \bar{\psi}^-\bar{\partial} \bar{\psi}^+ + \frac{3}{2}\bar{\partial}(\bar{\psi}^+\bar{\psi}^-) 
\end{align}
up to the BRST trivial terms. The topological BRST charges are given by
\begin{align}
Q_{\text{top}} = \oint G^+ = \oint \psi^+ j^- \ ,  \ \ \ \bar{Q}_{\text{top}} = \oint \bar{G}^- = \oint \bar{\psi}^- \bar{j}^+ \ .
\end{align}
Similarly, the twisted supercurrent is given by
\begin{align}
G^- = \psi^- \beta  \ , \ \ \ \bar{G}^+ = \bar{\psi}^+ \bar{\beta} \ . 
\end{align}

It is shown \cite{Mukhi:1993zb} that after taking the cohomology with respect to $Q_{\text{top}}$, $(\beta,\gamma)$ and $(B,C)$ will be decoupled and we are left with the $c=1$ variables. $\psi^-$ and $\bar{\psi}^+$ will be identified with the anti-ghost $b$ and $\bar{b}$, and $\psi^+$ and $\bar{\psi}^-$ will be identified with  the ghost $c$ and $\bar{c}$.

 The $c=1$ non-critical string theory is described by the world-sheet action (see \cite{Nakayama:2004vk} for a review)\footnote{Throughout this paper, we will use the convention $\alpha' = 2$.}:
\begin{align}
 \frac{1}{2\pi}\int d^2z \left( \partial X \bar{\partial}X + \partial \phi \bar{\partial}\phi + \frac{1}{\sqrt{2}} R \phi + 2\pi\mu e^{\sqrt{2} \phi}\right) \ \label{ncs}
\end{align}
together with the world-sheet reparametrization ghost $(b,c)$. Here $R$ is the world-sheet scalar curvature.
 We have used the same notation in the above coset construction to denote the fields appearing in the non-critical string theory so that the identification is obvious. We will compactify the $X$ boson at the self-dual radius $r=\sqrt{2}$.

\subsection{Operator mapping}
The claim is that the BRST cohomology of the A-twisted $SL(2,\mathbf{R})/U(1)$ Kazama-Suzuki coset at level $k=1$ is same as the physical observables of the $c=1$ non-critical string theory at the self-dual radius. We would like to refer \cite{Mukhi:1993zb}\cite{Ashok:2005xc} for the detailed analysis of the claim, and let us summarize the operator contents of the $c=1$ non-critical string theory at the self-dual radius. Once we admit that the cohomology agrees, the following list can be used as representatives of the observables in the A-twisted $SL(2,\mathbf{R})/U(1)$ coset model written in the Wakimoto representation.\footnote{In the following, the right-mover is suppressed. When we combine the left-mover and the right-mover, we have to take care of the fact that $\phi$ is non-compact while $X$ is compactified at the self-dual radius.}

The ghost number zero states in the $c=1$ string theory are known as the ground ring \cite{Witten:1992yj}. They are labelled by two (half-)integers $(s,n)$ and denoted as $O_{s,n}$, where $s\ge0, -s\le n \le s$. Correspondingly, we can construct paired states with ghost number 1: $a O_{s,n}$ by acting $a=c\gamma$. The ground ring is generated by $O_{1/2,1/2} = (cb+\frac{\sqrt{2}}{2}(\partial\phi - i\partial X)) e^{\frac{\sqrt{2}}{2} iX - \frac{\sqrt{2}}{2} \phi}$, and $O_{1/2,-1/2} = (cb+\frac{\sqrt{2}}{2}(\partial\phi + i\partial X)) e^{-\frac{\sqrt{2}}{2} iX - \frac{\sqrt{2}}{2} \phi}$.

Other important series of operators are tachyon operators that are given by 
\begin{align}
Y^{+}_{s,-s} = c e^{-i\sqrt{2}sX + \sqrt{2}(1-s)\phi} \ ,
\end{align}
with $s\ge 0$. They correspond to the $\mathcal{N}=2$ Liouville chiral primary operator and its generalization (see e.g. \cite{Nakayama:2004vk} for a review on the relation between the $SL(2,\mathbf{R})/U(1)$ Kazama-Suzuki coset theory and the $\mathcal{N}=2$ Liouville theory). 

We can also construct the discrete states: 
\begin{align}
Y^{-}_{s,n} = (K^{s-n}) c e^{i\sqrt{2}s X + \sqrt{2}(1+s)\phi} \ ,
\end{align}
and their partners $aY_{s,n}^{-}$, where $K = \oint \beta e^{-i\sqrt{2}X}$ is the $SU(2)$ lowering operator. They have ghost number $1$ and $2$ respectively.

The other half of the cohomology is the BPZ dual of all the above operators. Up to the ghost factor, they are obtained by the reflection of the Liouville factor: $\sqrt{2}(1+s) \phi \to \sqrt{2}(1-s)\phi$.

For later purposes, we observe a simple formula for the $U(1)$ $R$-charge assignment. From the gauging relation, the $R$-charge, the ghost number and the $U(1)_X$ momentum are all related by the linear relation:
\begin{align}
Q_R = Q_{\text{ghost}}-2Q_X \ . \label{rs}
\end{align}
The relation \eqref{rs} is only true for the physical observables that are invariant under the $U(1)$ gauging.

\subsection{Relation to conifold}
The topological string theory (B-model) on the deformed conifold is equivalent to the $c=1$ non-critical string theory at the self-dual radius \cite{Ghoshal:1995wm}. First of all, the ground ring structure of the $c=1$ string theory suggests that the deformed conifold would somehow appear as an emerging geometry. The direct computation of the higher genus partition function shows that these two indeed agree with each other.

The topological partition function can be computed as
\begin{align}
F &= \log Z = \int \frac{dt}{t} \frac{e^{-t\mu}}{(2\sin(\epsilon t/2))^2} \ \cr
 & = \frac{1}{2}\tilde{\mu}^2\log\tilde{\mu} -\frac{1}{12}\log \tilde{\mu} + \sum_{g=2}\frac{B_{2g}}{2g(2g-2)}\tilde{\mu}^{2-2g} \ ,
\end{align}
where $\tilde{\mu} = \mu/\epsilon = \mu/g_s$ is a renormalized cosmological constant that is identified with the deformation parameter of the conifold divided the topological string coupling constant.

An alternative way to see the duality is to realize that the $SL(2,\mathbf{R})/U(1)$ coset model at level $k=1$ can be formally realized as a Landau-Ginzburg model with the superpotential $W = \mu X^{-1} + X_1^2 + X_2^2 +X_3^2+X_4^2$. A simple application of the Calabi-Yau/Landau-Ginzburg correspondence gives us the defining equation of the deformed conifold
\begin{align}
-\mu x^{-1} = x_1^2 + x_2^2 + x_3^2 + x_4^2
\end{align}
embedded in the weighted projective space $\mathrm{WCP}^4_{-2,1,1,1,1}$. The deformed conifold is obtained by simply setting $x=-1$.

The operators corresponding to the discrete tachyon $Y_{s,\pm s}^+$ are related to the deformation of the asymptotic complex structure moduli of the deformed conifold. The scattering amplitudes in the both sides are governed by the $W_{\infty}$ algebra. It has a structure of the Toda hierarchy and the partition function, which can be computed from the matrix model, is regarded as a $\tau$-function \cite{Dijkgraaf:1992hk}\cite{Imbimbo:1995yv}\cite{Aganagic:2003qj}.

However, we observe that the corresponding operators in the $SL(2,\mathbf{R})/U(1)$ coset model possess $Q_R \neq 1$, except for the cosmological constant operator $c e^{\sqrt{2}\phi}$, so in the physical string theory, it seems that they do not give us conformal invariant perturbations. A related point is the $SU(2)\times SU(2)$ symmetry of the deformed conifold. The deformed conifold has a geometric $SO(4) \simeq SU(2) \times SU(2)$ symmetry, and so does the $c=1$ non-critical string theory at the self-dual radius. The $SU(2) \times SU(2)$ symmetry of the $SL(2,\mathbf{R})/U(1)$ Kazama-Suzuki coset model is not manifest. In fact, the coset theory itself does not possess such a symmetry, while the topological twisted theory does.

This might clarify the above-mentioned puzzle about the topological string theory interpretation of the tachyon operators. The tachyon operator with $s$ unit of the $X$-momentum has spin $s$ representation under $SU(2)$. One can always perform the $SU(2)$ rotation so that $Y^+_{s,0}$ are the ``tachyon operators" in the $c=1$ non-critical string theory. In the rotated basis, they do have $U(1)$ $R$-charge  $Q_R=1$ and now they do correspond to the conformal (non-normalizable) deformations of the untwisted  $SL(2,\mathbf{R})/U(1)$ coset theory. 

The $SU(2)\times SU(2)$ symmetry does not commute with the $U(1)$ R-symmetry, nor untwisted Virasoro generators of the physical $SL(2,\mathbf{R})/U(1)$ coset theory, but it seems crucial to realize this symmetry to guarantee the deformed conifold interpretation. It is almost obvious from the above non-compact Gepner model construction because the $SU(2)\times SU(2)$ symmetry does not act on $\mu X^{-1}$ superpotential, so the appearance of the symmetry after the topological twist is totally miraculous.
It would be interesting to see how this hidden symmetry is realized within the coset field theory beyond the discussions relied on the Wakimoto representation.

\section{Refined cigar and $\Omega$-deformation}
In \cite{Antoniadis:2010iq}, it was shown that a certain class of higher derivative F-terms in $\mathcal{N}=2$ compactification can be computed from the topological string amplitudes. It is schematically denoted as
\begin{align}
\mathcal{T}_{g,n} = \int d^4x F_{g,n}(\hat{\phi}_I,\bar{\hat{\phi}}_I) (R^-)^2 (T^{-})^{2g-2} (F^+_I)^{2n} + \cdots \ . 
\end{align}
Here $R^{-}$ is the anti-self-dual Riemann tensor and $T^{-}$ is the anti-self-dual graviphoton field strength and $F^{+}_I$ is the self-dual field strength associated with the vector multiplet whose lowest component is $\hat{\phi}_I$.

A direct computation in the type II string theory \cite{Antoniadis:2010iq} relates the higher derivative F-terms with the topological string amplitudes:
\begin{align}
F_{g,n} = \int_{\mathcal{M}_{(g,n)}} \langle \prod_{k=1}^{3g-3+n} |(\mu_k \cdot G^{-})|^2 \prod_{k=1}^{n} \int \Psi_{I_k} \prod_{l=1}^n \hat{\Psi}_{J_l}\rangle_{\text{top}} \ . \label{refined}
\end{align}
Here, $\Psi_I$ are (anti chiral, chiral) primary operators with the $U(1)$ $R$-charge $(-1,1)$ and dimension $(1,1)$, which will be integrated over the Riemann surface. The hatted operators are defined as $\hat{\Psi}_{J} = \oint dz \rho(z) \oint d\bar{z} \tilde{\rho}(\bar{z}) \Psi_J$, where $\rho$ is the unique left-moving operator with the charge $+3$ and dimension $0$. Thus, $\hat{\Psi}_J$ have $U(1)$ $R$-charge $(+2,-2)$ and dimension $(0,0)$, and they are in the twisted BRST cohomology. They are located at $n$ distinct punctures on the genus $g$ Riemann surface.  $\mu_k$ is the Beltrami-differential associated with the complex structure moduli space $\mathcal{M}_{(g,n)}$ for the $n$-punctured genus $g$ Riemann surfaces. The natural measure for the A-twist is given by
\begin{align}
|(\mu_k \cdot G^{-})|^2 = (\mu_k \cdot{G}^-)(\bar{\mu}_k\cdot \bar{G}^+)  \ .
\end{align}

When $n=0$, the amplitude corresponds to the partition function of the genus $g$ topological string theory, and it computes the graviphoton corrections to the $\mathcal{N}=2$ prepotential. It is important to observe that the amplitude $F_{g,n}$ for $n \ge 1$ is not holomorphic because the insertion $\int \Psi_{I_k}$ is not necessarily annihilated by the topological BRST charge.

To evaluate the refined topological string amplitude \eqref{refined} in the deformed conifold theory, or equivalently for the $SL(2,\mathbf{R})/U(1)$ Kazama-Suzuki coset at level $k=1$, we would like to use the map reviewed in the previous section. 
In the Wakimoto representation, which is intuitively related to the $c=1$ variables, the operators appearing in the refined topological string amplitude are given by
\begin{align}
\hat{\Psi} &= Y^+_{\frac{1}{2},-\frac{1}{2}}\bar{Y}^+_{\frac{1}{2},-\frac{1}{2}} = c\bar{c}e^{-\frac{i\sqrt{2}}{2} X + \frac{\sqrt{2}}{2}\phi }  \cr
\rho & = Y^{+}_{1,-1}  = c e^{-i\sqrt{2}X} \cr
\Psi & = e^{i\frac{\sqrt{2}}{2}X + \frac{\sqrt{2}}{2} \phi} \ ,
\end{align}
where the subscripts $I$, $J$ are dropped because the deformed conifold allows only one (normalizable) vector multiplet.

The identification is straightforward. First of all, $\rho$ is supposed to be the unique $U(1)$ $R$-charge $+3$ operator in the coset theory with the ghost number $1$. From the $U(1)$ $R$-charge assignment, it means that the $X$-momentum is $-1$. The reason why the simplest tachyon vertex (rather than higher $Y^+_{n,-1}$) is chosen is as follows: otherwise the operator would have non-zero Liouville momentum, and as a consequence, the decomposition between the left-mover and the right-mover, which is necessary to define $\rho$ and $\bar{\rho}$ separately, would be impossible. 

With the similar reason, we have chosen $\hat{\Psi}$ to be a tachyon operator, which has a direct interpretation in the deformed conifold geometry. Again the $U(1)$ $R$-charge assignment dictates that $\hat{\Psi}$ has $-1/2$ unit of $X$-momentum, and the simplest possibility is the tachyon field $Y^+_{\frac{1}{2},-\frac{1}{2}}$. It is also important to recognize that the choice guarantees that $\hat{\Psi}$ does not have an OPE singularity themselves in the computation of the refined topological string amplitude.
Once we have fixed $\rho$ and $\hat{\Psi}$, the remaining operator $\Psi$ is automatically determined (up to BRST trivial terms).

Therefore, the refined topological string amplitudes can be rewritten as
\begin{align}
F_{g,n} = \int_{\mathcal{M}_{(g,n)}} \langle \prod_{k=1}^{3g-3+n} |\mu_k \cdot b\beta)|^2 \prod_{k=1}^{n} \int e^{i\frac{\sqrt{2}}{2}X + \frac{\sqrt{2}}{2} \phi}  \prod_{l=1}^n c\bar{c} e^{-\frac{i\sqrt{2}}{2} X + \frac{\sqrt{2}}{2}\phi }   \rangle_{c=1} \ , \label{refinedc}
\end{align}
where the correlation function is evaluated by the $c=1$ non-critical string theory at the self-dual radius whose action is given in \eqref{ncs}. Note that $G^{-}$ is replaced by $b\beta$. The ghost number conservation as well as $X$ momentum conservation are all satisfied.
This is the first main result of this paper.

Without using the Wakimoto representation, we can keep track of the origin of the vertex operator in the $SL(2,\mathbf{R})/U(1)$ coset theory. We begin with the chiral primary operator $\Phi_{J=-1/2, m=1/2}$ of the $SL(2,\mathbf{R})/U(1)$ coset model that corresponds to the cosmological constant operator $ce^{\sqrt{2} \phi}$ in the $c=1$ theory, and we act $a\sim c\gamma = [G^{-}_{-1/2}]^{-1}$ operation. Then we perform the inverse of the spectral flow operation $U^{-1}$, which brings us to $Y^{+}_{\frac{1}{2},-\frac{1}{2}}$ that will be identified with $\hat{\Psi}$. By repeating the same procedure, we obtain $Y^{+}_{1,-1}$ which corresponds to $\rho$. In summary, we have the correspondence\footnote{For $\hat{\Psi}$ and $\Psi$, we will repeat the same procedure for the right-mover as well, which is implicit in \eqref{cosetv}.}:
\begin{align}
\hat{\Psi} &= U^{-1} a \Phi_{J=-1/2,m=1/2} \cr
\rho & = U^{-1}aU^{-1}a \Phi_{J=-1/2,m=1/2} \cr 
\Psi &= a^{-1} U  \ . \label{cosetv}
\end{align}
All these operations are intrinsic to the $SL(2,\mathbf{R})/U(1)$ coset theory without referring to the particular Wakimoto representation. Indeed, the correlation functions of the $c=1$ theory are reproduced from the $SL(2,\mathbf{R})/U(1)$ coset model \cite{Nakamura:2005sm} by using the Stoyanovsky-Ribault-Teschner relation \cite{Stoyanovsky:2000pg}\cite{Ribault:2005wp}.

The remaining task is to compare the refined topological amplitude \eqref{refinedc} with the radius deformation of the $c=1$ string theory. We would like to regard \eqref{refinedc} as deformed $c=1$ string theory amplitudes perturbed by the radius changing operator $\int d^2z \partial X \bar{\partial} X$ added to the $c=1$ action at the self-dual radius. 

Superficially, we do not see direct equivalence because the refined topological vertex is perturbed by $\int e^{i\frac{\sqrt{2}}{2}X + \frac{\sqrt{2}}{2} \phi}  \prod_{l=1}^n c \bar{c}e^{-\frac{i\sqrt{2}}{2} X + \frac{\sqrt{2}}{2}\phi }$ and not by $\int d^2z \partial X \bar{\partial} X$. However, it is important to realize, as argued in \cite{Antoniadis:2010iq}, that the refined topological amplitude \eqref{refined} or \eqref{refinedc} has a singularity when $\Psi$ approaches $\hat{\Psi}$, and the leading singularity $\sim \frac{1}{|z|^2}$ must be subtracted to obtain the 1PI effective action, and it is this 1PI effective action that will be compared with the $c=1$ radius deformation.

The leading singularity comes from the OPE of the $X$ boson:
\begin{align}
e^{i\frac{\sqrt{2}}{2}X}(z) \cdot e^{-i\frac{\sqrt{2}}{2}X}(w) \sim \frac{\mathbf{1}}{|z-w|^2} + \partial X \bar{\partial}X + \cdots 
\end{align}
where the overall coefficient (that will also depend on $|z-w|$) is omitted for simplicity. From the above argument, it is clear that one has to subtract the $1/|z-w|^2$ term in the evaluation of the 1PI effective action. This leaves us with the regularized deformation
\begin{align}
\int e^{i\frac{\sqrt{2}}{2}X + \frac{\sqrt{2}}{2} \phi}  \prod_{l=1}^n c\bar{c} e^{-\frac{i\sqrt{2}}{2} X + \frac{\sqrt{2}}{2}\phi } \to \int c\bar{c} \partial X \bar{\partial} X  + \cdots \ . \label{regu}
\end{align}
The exponent of the Liouville factor is determined so that the perturbation is conformally invariant.

The OPE of the Liouville part might require a little bit more attention. Usually, the OPE in the Liouville theory is defined in the physical Liouville momentum range: $e^{\sqrt{2}(1+ip)\phi}$, where $p\in \mathbf{R}$ (see e.g. \cite{Nakayama:2004vk} for a review). Here, we are discussing the unphysical region of the Liouville momentum $e^{\sqrt{2}(1-s)\phi}$, so the OPE must be regarded as an ``analytic continuation" of the three-point function defined in the physical range. However, the OPE in the unphysical range of the Liouville momentum formally makes sense, and indeed it was even used in the original derivation of the three-point function \cite{Dorn:1994xn}\cite{Zamolodchikov:1995aa}\cite{Teschner:1995yf}. We note that the OPE used in \eqref{regu} is precisely when the Liouville momentum is saturated by the cosmological constant operator insertion in the perturbative Liouville computation as done in \cite{Dorn:1994xn}\cite{Zamolodchikov:1995aa}.\footnote{A better approach might be to perform the double Wick rotation $e^{i\sqrt{2}sX} \to e^{\sqrt{2}sX} = e^{-i\sqrt{2}sT}$, and $e^{\sqrt{2}(1-s)\phi} \to e^{\sqrt{2}(1+is)\phi}$ in the scattering amplitude. The $c=1$ amplitudes are originally defined in these variables, and the OPE in the Liouville sector perfectly makes sense in this physical region.}

Can we say anything about the higher derivative deformations in \eqref{regu}? We know that the vertex $ e^{\pm i\frac{\sqrt{2}}{2}X + \frac{\sqrt{2}}{2} \phi}$ transforms as spin $\frac{1}{2}$ representation under the enhanced $SU(2)$ symmetry at the self-dual radius. As long as the higher derivative terms that could appear in \eqref{regu} are within the observables of the $c=1$ theory, they cannot appear because the higher derivative terms are in the higher spin representation than $1$. The only possible term is the radius changing operator that lies in the spin $1$ representation. We conclude that the regularized 1PI action can be computed by perturbing the $c=1$ non-critical string amplitude by a radius changing operator.

We note that after subtracting the leading order singularity, the computation of the refined topological string amplitude {\it is} holomorphic in our example (up to a possible holomorphic anomaly) because $c\bar{c}\partial X\bar{\partial}X$ is in the BRST cohomology of the $c=1$ string theory denoted by $Y_{(1,0)}^+$ (and hence in the topologically twisted theory). Similarly, from the charge assignment, if one subtracts the leading order singularity, the first non-trivial OPE between $\Psi_I$ and $\hat{\Psi}_J$ should give birth to the operator whose $R$-charge is $(1,1)$ and the twisted conformal dimension $(0,0)$. Unless it is given by a certain descendent operator, it has the same charge as the topological chiral primary operator, and the amplitude is likely to be holomorphic (up to a possible holomorphic anomaly).

Finally, to see the connection to the heterotic string computation \cite{Morales:1996bp}, we introduce the generating function for the (regularized) amplitudes 
\begin{align}
F(\epsilon_-, \epsilon_+) = \sum_{g=1}^\infty \sum_{n=1}^\infty \epsilon_-^{2g} \epsilon_+^{2n} F_{g,n} \ .
\end{align}
The heterotic partition function 
\begin{align}
F \sim \int \frac{dt}{t} \frac{\pi \epsilon_1}{\sin(\pi \epsilon_1 t)} \frac{\pi \epsilon_2}{\sin(\pi \epsilon_2 t)} e^{-t \mu} \ , \label{opat}
\end{align}
where $\epsilon_1 = \epsilon_+ + \epsilon_-$ and $\epsilon_2 = \epsilon_+ -\epsilon_-$, is nothing but the free energy of the $c=1$ non-critical string theory at the radius $\sqrt{2}|\epsilon_1/\epsilon_2|$ \cite{Gross:1990ub}. The change of the radius is induced by a small unit $\frac{\epsilon_+}{\epsilon_-} \sim \mu \epsilon_+$. 

To reproduce the $\mu$ dependence from our computation, we first normalize $\Psi$ and $\hat{\Psi}$ as $\Psi = \sqrt{\mu} e^{\frac{i\sqrt{2}}{2} X + \frac{\sqrt{2}}{2}\phi }$ and $\hat{\Psi} = \sqrt{\mu} c e^{-\frac{i\sqrt{2}}{2} X + \frac{\sqrt{2}}{2}\phi }$. The normalization by the factor $\sqrt{\mu}$ is chosen so that the insertion does not introduce any power of $\mu$ from the Liouville correlation function because otherwise the relation between the genus expansion and the $1/\mu$ expansion is lost. Then, on the right hand side of \eqref{regu}, we will gain no powers of $\mu$ after taking care of the Liouville OPE. Finally, we note that the $|\beta b|^2$ in the Beltrami differential is replaced by $\mu b \bar{b}$. The rule $\beta \to \sqrt{\mu}$ in the twisted cigar was first advocated in \cite{Mukhi:1993zb}. After this substitution, we see that all the factors combine so that our refined topological amplitude correctly reproduce the radius change by the amount $\mu \epsilon_+$ as in the heterotic string computation. 

We note that the $\Omega$-deformed partition function \eqref{opat} is closely related to the special function known as the Barnes double Gamma function $\Gamma_b(x)$ (see e.g. \cite{Nakayama:2004vk} for its property and its usage in the Liouville theory) upon the analytic continuation in $\epsilon_1$ and $\epsilon_2$ and subtracting the singularity: $F(\mu,\epsilon_1,\epsilon_2) \sim \log \Gamma_b(\mu +\frac{b+b^{-1}}{2})$, where $b= |\epsilon_1/\epsilon_2|$. The Barnes double Gamma function has a manifest ``T-duality" property $b \to b^{-1}$, and it is ubiquitous in the computation of the Liouville correlation functions with the central charge $c= 1+6(b+b^{-1})^2$. The double appearance of the double Gamma function for the radius deformation of the $c=1$ non-critical string theory as well as the change of the background charge in the Liouville field theory manifests the duality philosophy to connect the world-sheet formulation (radius deformation) and the target space Kodaira-Spencer theory formulation over the Riemann surfaces (changing of the background charge) advocated in \cite{Dijkgraaf:2009pc}.

\section{Discussion}
In this paper, we have studied the refined topological amplitudes for the deformed conifold by using the $SL(2,\mathbf{R})/U(1)$ Kazama-Suzuki coset construction. After subtracting the leading singularity to procure the 1PI effective action, we have shown that the refinement corresponds to the radius deformation of the $c=1$ non-critical string theory at the self-dual radius as proposed in the literature.

We note that the $c=1$ non-critical string theory at the self-dual radius has an enhanced $SU(2)$ symmetry. The refinement given by the radius deformation transforms as spin 1 representation of the $SU(2)$. More precisely, they form a triplet 
\begin{align}
Y_{1,1}^+ &= -ce^{i\sqrt{2}X} \cr
Y_{1,0}^+ &= c\sqrt{2}i\partial X \cr
Y_{1,-1}^+ & = c e^{-i\sqrt{2} X}  \ ,
\end{align}
and similarly for the right-mover.
Therefore, we expect an $SU(2) \times SU(2)$ rotated (but physically equivalent) version of the refinement. It would be interesting to find out its significance in the topological string theory, or $\mathcal{N}=2$ gauge theory.

We also note that more complicated non-compact Calabi-Yau space can be generated by using the non-compact Gepner-model construction with the usage of the $SL(2,\mathbf{R})/U(1)$ Kazama-Suzuki coset model with $k\neq 1$. It was conjectured that they are related to the $c<1$ non-critical string theory after twisting as in the deformed conifold case \cite{Sahakyan:2005dh}\cite{Niarchos:2005ny}\cite{Ashok:2006vz}. Then, we naturally expect that the refined topological amplitudes might have similar $c<1$ non-critical string theory interpretations.

Finally, the recent developments in non-rational CFTs have enabled us to compute various quantities directly and exactly within the CFTs. The above mentioned triality is now understood beyond the cohomology of the observables, but at the correlation function levels. It would be important to generalize the relation further to take into account the refinement discussed in this paper.

\section*{Acknowledgements}
The author would like to thank C.~Beem for stimulating discussions.
The work was supported in part by the National Science Foundation under Grant No.\ PHY05-55662 and the UC Berkeley Center for Theoretical Physics. 

\appendix

\end{document}